\documentclass[conference]{IEEEtran}
\IEEEoverridecommandlockouts
\usepackage{cite}
\usepackage{amsmath,amssymb,amsfonts}
\usepackage{algorithmic}
\usepackage{graphicx}
\usepackage{textcomp}
\usepackage{xcolor}
\usepackage{mathdots}
\usepackage{caption}
\usepackage{subfigure}
\def\BibTeX{{\rm B\kern-.05em{\sc i\kern-.025em b}\kern-.08em
    T\kern-.1667em\lower.7ex\hbox{E}\kern-.125emX}}

\begin{document}

\title{Universal Polar Coding for Parallel Gaussian Channels with Non-Binary Inputs and Its Applications to {HARQ} and {MIMO}\\
}

%\author{\IEEEauthorblockN{David Tse}
%\IEEEauthorblockA{\textit{Dept. of Electrical Engineering} \\
%\textit{Stanford University}\\
%California, USA \\
%dntse@stanford.edu}
%\and
%\IEEEauthorblockN{Bin Li}
%\IEEEauthorblockA{\textit{Dept. Communication Technology Research} \\
%\textit{Huawei Technologies}\\
%Shenzhen, P. R. China \\
%binli@huawei.com}
%\and
%\IEEEauthorblockN{Kai Chen}
%\IEEEauthorblockA{\textit{Dept. Communication Technology Research} \\
%\textit{Huawei Technologies}\\
%Shenzhen, P. R. China \\
%kaichen@ieee.org}
%\and
%\and
%\and

\author{\IEEEauthorblockN{Bin Li,  Jiaqi Gu and Huazi Zhang}
\IEEEauthorblockA{Wireless Technology Lab, Huawei Technologies, Shenzhen, P. R. China\\
Email: \{binli.binli,gujiaqi,zhanghuazi\}@huawei.com}}
\maketitle

\begin{abstract}
In this paper, we first propose an universal polar coding scheme for parallel Gaussian channels with non-binary inputs. It is assumed that the encoder knows only the sum capacity of M parallel channels instead of the capacity of any single channel. By decomposing each parallel channel into $T=\lceil \log_2 r \rceil$ sub channels, we therefore obtain {MT} binary sub-channels. A super polar coding scheme that across all sub-channels is then proposed. This scheme can achieve the sum capacity when the block length is sufficiently large. We have also discussed the applications of parallel polar coding design for both the {HARQ} and {MIMO} systems. It is shown that a capacity-achieving {HARQ} scheme can be obtained for block fading channel and a capacity-achieving {MIMO} design that requires only the feedback of the sum rate of all {MIMO} layers can also be attained.
\end{abstract}


\section{Introduction}
Polar codes are a major breakthrough in coding theory \cite{b1}. They are the first class of error correcting codes that can achieve Shannon capacity with a simple encoder and a simple successive cancellation decoder when the code block size is sufficiently large. Despite their wide applications in error correction field, recent researches have also illustrated multiple possibilities to extend the usage of polar codes into various communication scenarios such as source coding \cite{b2}, asymmetric channels \cite{b3}, BICM channels \cite{b4}, wiretap channels \cite{b5}, broadcast channels \cite{b6}, multiple access channels \cite{b7} and quantum channels \cite{b8}. In this paper, we consider the application of a super polar coding scheme in parallel Gaussian channels and its extensions in various scenarios.

Polar codes have a very nice (nesting) property that leads to the reliability ordering of polarized channels, namely, a polar code with lower coding rate is a subset of another polar code with higher coding rate if the corresponding two channels are sequentially degraded. Using this property, a universal polar coding scheme \cite{b9} was proposed for the parallel Gaussian channels that consisting of M independent BIAWGN channels. This scheme is proved to be capacity achieving with knowing only the sum rate in advance at the transmitter side rather than the capacity of any channel.

In this paper, we expand and extend the previous design to parallel Gaussian channels with non-binary inputs and consequently, obtain a more generalized polar coding approach. After that, we apply this super polar coding technique to both the HARQ and MIMO systems and accordingly demonstrate that, the proposed application in HARQ scenario is optimal and can achieve the capacity of the block fading channels. Furthermore, when extend the proposed design to MIMO systems, we show that the sum capacity can also be achieved with only feedback of the information of sum rate.

This paper is organized as follows. In section II, we introduce the basics of polar codes and its nesting property, in section III, we propose a new super polar coding scheme for parallel Gaussian channels with non-binary inputs. In section IV, we propose a polar coding scheme for HARQ system. The application of the proposed polar coding scheme in MIMO system is further described in section V. Finally, brief conclusions are draw in section VI.

\section{Polar code and its nesting property}
\subsection{Basics of Polar Codes}
Consider a BMS channel $W(y|x)$ with binary input alphabet $\mathcal{X}$ and output alphabet $\mathcal{Y} \subseteq \mathbb{R}$, polar codes are block codes with length $N=2^n$ for all integers $n \geq 0$. Let $G_N=B_NF^{\otimes n}$ be the generator matrix where $B_N$ is a bit-reversal permutation matrix, $F=\begin{bmatrix}1 & 0\\1&1\end{bmatrix}$ refers to the base matrix or kernel of the polar codes, $F^{\otimes n}=F \otimes F^{\otimes n-1}$ is a $N \times N$ matrix, $\otimes n$ denotes $n$-th Kronecker power. Let the $n$-bit binary representation of integer $i$ be $b_1,b_2,...,b_n$. The $n$-bit representation $b_n,b_{n-1},...,b_1$ is a bit-reversal order of $i$. The polar code is generated by
\begin{eqnarray}
x_1^N=u_1^NG_N=u_1^NB_NF^{\otimes n}
\end{eqnarray}
where $x_1^N=(x_1,x_2,...,x_N)$ is the encoded bit sequence, and $u_1^N=(u_1,u_2,...,u_N)$ is the encoding bit sequence.
The construction of polar code is based on the channel polarization phenomenon introduced in \cite{b1}, by combining and splitting the channels, the $K$ most reliable synthetic ones are used for transmitting the information bits and their indices form the information set $\mathcal{I}$, the complementary set  $\mathcal{F}$ is referred to the frozen set and set to zero for simplicity. Let $S$ represent a subset containing the information bit indices, then the coding rate is given by $R=|S|/N$, where $|S|$ is the size of  $S$.

\subsection{Nesting Property of Polar Codes}
Consider two BMS channels  $W$ and $V$ with a common input alphabet $\mathcal{X}$ and two output alphabets $\mathcal{Y}_1$ and $\mathcal{Y}_2$, and the transition probability per each channel is $p_1(y_1|x)$ and $p_2(y_2|x)$ respectively. In this case, we define that channel $V$ is stochastically degraded with respect to channel $W$  if there exists an intermediate channel $W'$ such that $W-W'-V$ forms a Markov chain, with input alphabets given by $\mathcal{Y}_1$, output alphabets given by $\mathcal{Y}_2$ and the transition probability $p_3(y_2|y_1)$ satisfy the following equation:
\begin{eqnarray}
p_2(y_2|x)=\sum_{y_1 \in \mathcal{Y}_1} p_1(y_1|x)p_3(y_2|y_1)
\end{eqnarray}
for all $x \in \mathcal{X}$ and $\mathcal{Y}_2 \in \mathcal{Y}$, denoted by $V \preceq W$.
Let $\mathcal{I}_V$ and $\mathcal{I}_W$ be the information set of polar codes designed for channel $V$ and $W$, respectively. According to the nesting property of polar codes that polarization operations preserve the degradedness \cite{b10} , we have $\mathcal{I}_V \subseteq \mathcal{I}_W$ when $V \preceq W$. Thus, consider a series of $Q$ sequentially degraded BMS channels: $W_1,W_2,...,W_m,...,W_Q$ where $W_1\preceq W_2...\preceq W_m \preceq ...\preceq W_Q$, with channel capacity $r_m=mR/Q, 1 \leq m \leq Q$, respectively. Let $\mathcal{I}_m$ be the information set of polar codes designed for channel $W_m$ with block length $N$. Let $K= \underset{1\leq m \leq Q}{\min}  \{|\mathcal{I}_m|\frac{Q}{m}\}$. We can find $Q$ disjoint subsets $S_1,S_2,...,S_Q$ of $[N]$, such that $\bigcup_{i=1}^m S_i \subseteq \mathcal{I}_m$. Namely, we can construct rate-compatible polar codes \cite{b9} based on nesting property as follows: initially, divide $K$ information bits into $Q$ subsets as $S_1,S_2,...,S_Q$, respectively, where $S_1$ contains the indices of the top $K/Q$ most reliable bits, $S_2$ contains the indices of the second top $K/Q$ most reliable bits,..., and $S_Q$ contains the indices of the least reliable $K/Q$ bits. Accordingly, $Q$ capacity-achieving polar codes with rates $R/Q,2R/Q,...,$ and $R$  can be obtained, respectively, and they are rate-compatible as $C_1 \subset C_2 \subset ... \subset C_Q$, where $C_1,C_2,...,C_Q$ are $Q$ polar codes.

\section{Super Polar Coding for Parallel Gaussian Channels with Non-Binary Inputs}

\begin{figure}[h]
\centering
  \includegraphics[width=8.5cm,height=6cm]{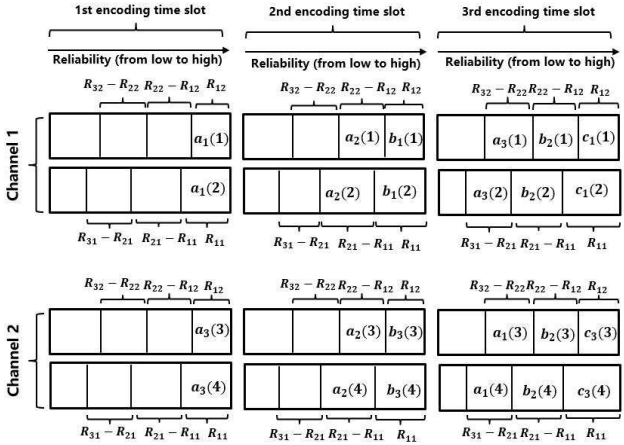}
  \caption{Polar coding for M=2 parallel Gaussian channels, Q=3, T=2}
  \label{fig2}
\end{figure}

We have proved in \cite{b9} that with knowing only the sum capacity of the parallel binary Gaussian channels, our proposed design can achieving the channel capacity. In this section, we extend the results into the non-binary input scenario and use a simple example in Fig. \ref{fig2} to illustrate that.

We illustrate the general structure of our proposed design as follows. Suppose that we have $M$ parallel channels with rates $r_1, r_2,..., r_M$ respectively, and the sum capacity is denoted by $r(r>1)$ where $r=r_1+r_2+...+r_M$. Each parallel channel is further decomposed into $T$ binary sub-channels, where $T=\lceil \log_2 r \rceil$  and $\lceil x \rceil$ is the least integer which is not less than $x$, i.e., $\lceil x \rceil \geq x$. Let $r_k= m_k r/Q$, where $m_k$ and $Q$ are integers, $1\leq k \leq M$, $m_1+m_2+...+m_M=Q$. Note that when $Q$ is sufficiently large, $m_k r/Q$ can closely approximate any arbitrary value $r_k (0\leq r_k\leq r)$. Assume that a parallel channel with capacity $q r/Q$ ($q$ is integer and $0\leq q \leq Q$) is decomposed into $T$ binary sub-channels, the corresponding sub-rate per each sub-channel $r_{q,1},r_{q,2},...,r_{q,T}$ satisfy that $q r/Q=r_{q,1}+r_{q,2}+...+r_{q,T}$.

When compared with the parallel polar coding scheme for binary input channel that we have proposed in \cite{b9}, the difference is that there are $T$ polar encoders instead of one polar encoder for each parallel channel, and the encoding bits used for one polar encoder in \cite{b9} are spread over $T$ polar encoders in the non-binary input scenario.

Specifically, all $K$ non-binary message bits are first divided into $Q$ sub-blocks, i.e, $(a_1,a_2,...,a_Q)$. As stated before, for each parallel channels, there are $T$ polar encoders equipped for the decomposed $T$ binary sub-channels. The upcoming information blocks denoted by $(b_1,b_2,...,b_Q)$, $(c_1,c_2,...,c_Q)$,..., are processed in the same way. Namely, each input $K$ information bits are split, encoded, mapped and then transmitted over each non-binary parallel channel. Specifically, we allocate the information bits via different ordering pattern for each parallel channels.

We summarize the detailed process as follows:
\begin{itemize}
\item The $K$ input bits are divided into $Q$ bit sequence, namely, $(a_1,a_2,...,a_Q)$. These $Q$ sub-blocks are distributed over $Q$ encoding time-slots. The incoming block is then divided
      into $Q$ sub-blocks and we obtain $(b_1,b_2,...,b_Q)$, thus, the second block is distributed over the 2rd to the $(Q+1)$-th encoding time slots consequently. Similarly, the incoming blocks with $K$ information bits are allocated to the corresponding time slots with the same manner.\vspace{0.2cm}

\item For the first parallel channel, $(a_1,a_2,...,a_Q)$ is distributed over the first $Q$ encoding time slots. $a_1$ is further divided into $T$ sub-blocks and placed into the most reliable positions in the first encoding time slot with $N R_{1,1},N R_{1,2},...N R_{1,T}$ bits, respectively. $R_{1,t}$ is the code rate of the first block at the $t$-th ($1\leq t \leq T$) sub-channel. Consequently, for the $q$-th sub-block $a_q$, where $(2\leq q \leq Q)$ , it is also distributed over $T$ polar encoders and placed in the $q$-th most reliable positions in the $q$-th encoding time slot, with $N(R_{q,t}-R_{q-1,t})$ bits in the $T$ polar encoders, respectively. Note that each sub-block $a_q$ contains the same number of bits $r/Q$. The following blocks are then distributed over corresponding encoding time slots in a same way. \vspace{0.2cm}

\item For the second parallel channel, the encoding sub-blocks are the reversal order of $(a_1,a_2,...,a_Q)$. When the amount of parallel channels increases to three, for the third parallel channel, the encoding sub-blocks are transformation of $(a_1,a_2,...,a_Q)$ using a polar matrix \cite{b9}. When the number of the parallel channels is $M$ $(M>3)$, the encoding sub-blocks for each parallel channel are transformation of $(a_1,a_2,...,a_Q)$ using one of the $M$ universal decodable matrices (UDM) \cite{b9} \cite{b11}.\vspace{0.2cm}

\end{itemize}

Fig. \ref{fig2} shows sub-block distribution for $M=2$ parallel channels with $T=2$ and $Q=3$. Let $(a_1,a_2,a_3)$ be the first information block which has three sub-blocks. These three sub-blocks $a_1$, $a_2$ and $a_3$ are distributed over the first three encoding time-slots, and each sub-block is further distributed over $T=2$ polar encoders corresponding to $T=2$ sub-channels. When $a_1$ is encoded for the first parallel channel, it is divided into two sub-blocks: $a_1 (1)$ and $a_1 (2)$, which contains $NR_{1,1}$ and $NR_{1,2}$ information bits, respectively. While when $a_1$ is encoded for the second parallel channel, it is divided into two sub-blocks: $a_1 (3)$ and $a_1 (4)$, which contains $NR_{3,1}-NR_{2,1}$ and $NR_{3,2}-NR_{2,2}$ information bits, respectively. Each sub-block is divided differently for T different parallel channels. For the second channel transmission, we perform the same processing but using the reversal order, namely, $(a_3,a_2,a_1)$ is distribute and encoded. Other information blocks which also has three sub-blocks are distributed over corresponding encoding time slots, respectively as shown by fig. \ref{fig2}. Suppose that the first parallel channel has a rate $q r/Q, (0\leq q \leq Q)$, then the second parallel channel has a rate $(Q-q)r/Q$. We can decode $q$ sub-blocks of $(a_1,a_2,a_3)$ in the first $q$ encoding time-slots from the first channel, and decode $Q-q$ sub-blocks of $(a_1,a_2,a_3)$ in the first $Q-q$ encoding time-slots from the second channel. Then we can obtain all sub-blocks of $(a_1,a_2,a_3)$. After cancellation of $(a_1,a_2,a_3)$, we can decode the second block $(b_1,b_2,b_3)$, and so on, until we decode all information blocks. Similarly, this also applies to the other scenarios when $M \geq 3$.

\section{Application I: An Optimal HARQ Scheme For Block Fading Channel}
In this section, using the polar coding scheme for parallel Gaussian channel, we propose a capacity-achieving polar coding HARQ scheme for block fading channel. For any arbitrary number of transmission, assuming that channels remain constant during each transmission but are independent from each other.

\vspace{-0.1cm}
\begin{figure}[h]
\centering
  \includegraphics[width=8cm,height=4cm]{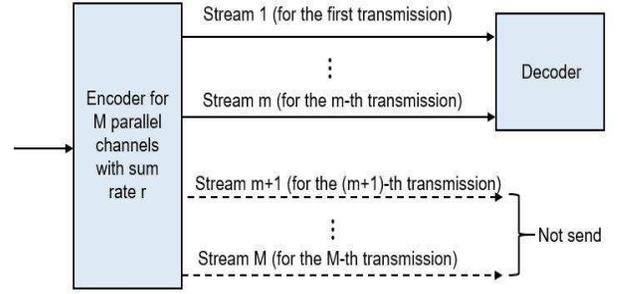}
  \caption{An optimal HARQ scheme for W transmissions}
  \label{fig3}
\end{figure}

An ideal and capacity-achieving HARQ scheme can be formed as follows. Suppose that a code block with rate $R$ is sent in the first transmission, if channel capacity $r_1 \geq R$, then this packet can be correctly decoded, otherwise the second transmission is carried out. Let the channel capacity of the second transmission be $r_2$, if  $r_1+r_2 \geq R$, then the information bits can be decoded with vanishing error probability. Accordingly, for any $M$ transmissions, once the sum rate satisfies $r_1+r_2+...+r_M \geq R$, the receiver can then jointly and successfully decode the corresponding packets. As denoted by Fig. \ref{fig3}, this is equivalent that we have $M$ parallel channels with the first $m$ channels having a sum rate larger than or equal to $R$ and the rest $M-m$ channels have rates 0s. The overall sum rate is larger than or equal to $R$. Therefore we can successfully decode all information bits.

\section{Application II: MIMO System with Simple Sum Rate Feedback}
\begin{figure}[h]
\centering
  \includegraphics[width=6.75cm,height=3.75cm]{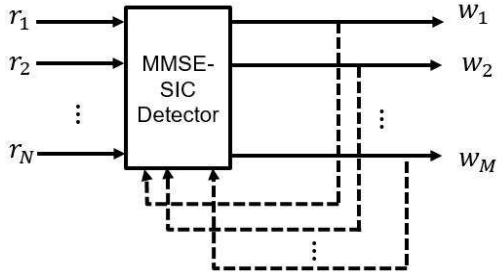}
  \caption{MMSE-SIC detection of MIMO}
  \label{fig6}
\end{figure}

In this section, we consider the MIMO scenario which is widely used in wireless communications. The capacity of the MIMO system can be illustrate as $C=\log_2 \textrm{det}(I_r + \textrm{SNR}\times HH^H)$, where $\textrm{det}(x)$ is the determinant of matrix $x$, $I_r$ is an unit matrix and $H$ is the channel matrix, $H^H$ is the Hermitian transpose of $H$. Obviously, the capacity can be significantly increased when compared with non-MIMO case. It is shown in \cite{b12} that the capacity can be achieved under the MMSE-SIC detection and the corresponding achievale sum-rate is given by $\sum_{i=1}^{N_r} \log_2 (1+ \gamma_i^{\textrm{MMSE-SIC}})$, where $N_r$  is the number of information layers which is not larger than the number of  receiver antennas, $\gamma_i$ is the SNR at $i$-th layer of MMSE-SIC detector. Fig. \ref{fig6} shows the MMSE-SIC detector and $\gamma_i$ is the SNR of detector output $w_i$. The MIMO capacity is the same as the sum rate of all layers under MMSE-SIC detection. In order to achieve the MIMO capacity, the receiver needs to feedback $\gamma_i$ of all layers or the rates of all layers. The transmitter can then allocate the exact rates at different layers.

\begin{figure}[h]
\centering
  \includegraphics[width=8cm,height=5cm]{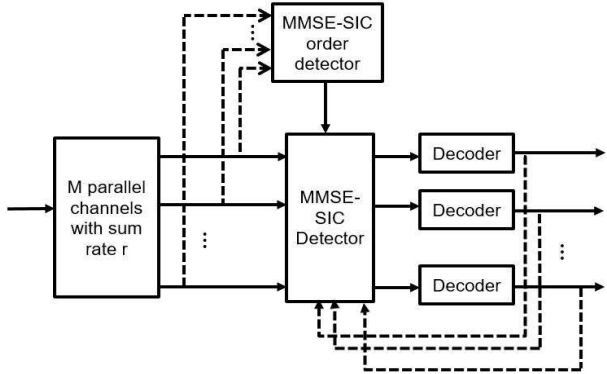}
  \caption{Capacity-achieving MIMO system with simple sum rate feedback}
  \label{fig7}
\end{figure}

Suppose that the MIMO system has $M$ layers and has a capacity rate $r$. Then we encode all information bits in the same fashion as for $M$ parallel Gaussian channels with a sum rate $r$. We obtain $M$ coded bit streams and transmit them over the $M$ layers of MIMO system as shown in Fig. \ref{fig7}.  Compared with the traditional feedback, our scheme only need a feedback of sum rate while the traditional scheme needs feedback of all rates of all layers.  This will reduce a lot of feedback overhead especially for large scale MIMO system.

\vspace{-0.1cm}

In order to achieve the MIMO capacity, the MMSE-SIC detector assumes that every layer can be successfully decoded and be completely cancelled from the received signals, and the following detection layers have no interference from the previously detected layers. Note that the proposed scheme requires a detection order according to the descending SNR order. This is due to the fact that when $M$ MIMO layers have rates  $q_1r/Q,q_2r/Q,...,q_Mr/Q$, respectively, where $q_m$ is an integer and $0 \leq q_m \leq Q, 1\leq m\leq M$. The  $m$-th layer can decode all information bits in the first $q_m$ encoding time slot or the first $q_mT$ polar codes. That is to say, if the $m_1$-th layer is decode before the $m_2$-th layer $(q_{m_1}< q_{m_2})$, then at the $m_1$-th decoding stage, we can obtain the decoded information bits in the first $q_{m_1}$ encoding time-slots while at the $m_2$-th layer, information bits in the first $q_{m_2}$ encoding time slot can be obtained. Which, indicates that the $m_2$-th layer will decode the message bits in the  $q_{m_2}-q_{m_1}$ encoding time-slots containing interference from the $m_1$-th layer. Therefore, current decoding layer would not interfere subsequent layers if we follow the SNR descending order for decoding.

\vspace{-0.1cm}

\section{Conclusions}
In this paper, we propose a super and universal polar coding scheme for $M$ parallel Gaussian channels with non-binary inputs. We decompose each parallel channel into T binary sub-channels so that $MT$ binary sub-channels are obtained. We have proved in \cite{b9} that with knowing only the sum capacity of the parallel channels, we can realize the capacity-achieving transmission. Thus, we introduce this results into the non-binary input scenario and bring the corresponding design.
After that, we extend the idea to both HARQ and MIMO systems. For HARQ system, the proposed scheme is optimal and capacity-achieving for block fading channels. Specifically, for any arbitrary $M$ transmissions, the channels remain constant during each transmission but are independent from each other. Suppose that the information rate is $R$ in the first transmission, and the channel capacity rates are $r_1$,$r_2$,..., and $r_M$  for the $M$ transmissions, respectively, once the sum rate satisfies $r_1+r_2+...+r_M \geq R$, the information bits can be decoded with vanishing error probability. As for the applications in MIMO system, the proposed technique can achieve MIMO capacity with knowing only the feedback of sum rate, compared to the traditional feedback of all rates of all layers, significant overhead can be reduced especially for large scale MIMO scenarios.

\end{document}